\documentclass[%
 reprint,
 amsmath,amssymb,
 aps,
 prl,
 superscriptaddress,
 longbibliography
]{revtex4-2}

\usepackage{comment}
\usepackage[T1]{fontenc}
\usepackage{lmodern}
\usepackage[utf8]{inputenc}
\usepackage{bm}
\usepackage{physics}
\usepackage{booktabs}  
\usepackage{microtype}
\usepackage{graphicx}
\usepackage{xcolor}
\usepackage[normalem]{ulem}
\usepackage{comment}
\usepackage{siunitx}
\usepackage[version=4]{mhchem}

\usepackage{hyperref}
\hypersetup{
    bookmarks=true,          % Show bookmarks bar
    unicode=true,            % Allow unicode in bookmarks
    pdftoolbar=true,         % Show Acrobat's toolbar
    pdfmenubar=true,         % Show Acrobat's menu
    pdffitwindow=false,      % Page fit to window when opened
    pdfstartview={FitH},     % Fit width of page to window
    pdftitle={Polaritonic rotor cat states},    % PDF title
    pdfauthor={Volker Karle},   % PDF author
    colorlinks=true,         % Colored links instead of boxes
    linkcolor=blue,          % Color of internal links
    citecolor=blue,          % Color of citation links
    filecolor=magenta,       % Color of file links
    urlcolor=magenta         % Color of external links
}

\graphicspath{{./figures/}}
\renewcommand{\and}{\quad \text{and}\quad}

\begin{document}

\title{Collective rotational cat states of molecules in microwave cavities}

\author{Volker Karle}
\email{vkarle@ist.ac.at} % corresponding author
\thanks{These authors contributed equally.}
\author{Florian Kluibenschedl}
\thanks{These authors contributed equally.}
\author{Mikhail Lemeshko}
\affiliation{Institute of Science and Technology Austria (ISTA), Am Campus 1, 3400 Klosterneuburg, Austria}
\author{Vasil Rokaj}
\email{vasil.rokaj@villanova.edu}
\affiliation{Department of Physics, Villanova University, Villanova, Pennsylvania 19085, USA}

\begin{abstract}
We show theoretically that an ensemble of polar molecules coupled to a microwave cavity supports hybrid rotational–photonic cat states. The cavity couples to a symmetric rotor in the bright manifold of $N$ molecules with $\sqrt{N}$-enhancement. In the dispersive limit of the collective strong coupling regime, virtual multilevel transitions induce an effective Kerr nonlinearity, as confirmed by Wigner tomography and a Schrieffer-Wolff analysis, leading to parity-locked cat structure in the cavity sectors. Collective molecular rotations thus provide a new route to hybrid light–matter cat states.
\end{abstract}

\maketitle

\textit{Introduction.} Schr{\" o}dinger cat states, nonclassical superpositions of macroscopically distinct coherent states, are key resources for quantum computing, communication, and sensing~\cite{Mirrahimi2014,Ralph2003PRA,Gilchrist2004JOSAB,heFastGenerationSchrodinger2023}. Their preparation and stabilization, however, remain challenging. Existing approaches include probabilistic measurement-based protocols~\cite{Ourjoumtsev2006Nature,Deléglise2008Nature}, engineered dissipation with multi-tone control in circuit QED~\cite{Leghtas2015Science,Touzard2018PRX,Grimm2020Nature,Lescanne2020NatPhys,hajrHighCoherenceKerrCatQubit2024}, and precisely timed unitary synthesis~\cite{Kienzler2016Science,Vlastakis2013Science,Sun2014Nature}. Each route comes with characteristic limitations: non-determinism in measurement-based schemes, substantial control overhead in dissipative stabilization, and the transient nature of purely unitary protocols.

Molecules offer a complementary route in which fabricated nonlinear elements can be replaced by intrinsic vibrational and rotational anharmonicity~\cite{Albert2020Molecules,Bohn2017Science,RevModPhys.91.035005,Lemeshko_chapter,PhysRevLett.130.103202}. Recent ultracold-molecule experiments have established rotational motion as a coherent and interacting quantum resource by demonstrating long-lived rotational qubits, dipolar entanglement, and two-qubit gates in optical tweezer arrays~\cite{burcheskyRotationalCoherenceTimes2021,gregorySecondscaleRotationalCoherence2024,picardEntanglementISWAPGate2025,ruttleyLonglivedEntanglementMolecules2025,baoDipolarSpinexchangeEntanglement2023}. 

At the same time, polaritonic chemistry has emerged as a novel approach to control chemical reactivity and molecular dynamics~\cite{EbbesenRubioScholes, Ebbesen2016}. Experiments have suggested the control of molecular dynamics through the coherent collective strong coupling in large molecular ensembles in microcavities~\cite{Hutchison2012, Hutchison2013, Lather2019, SimpkinsScience, EbbesenNMR, Murakoshi},
finding that optimal modifications to the dynamics occur when
the cavity is in resonance with molecular excitations~\cite{Hutchison2012, Hutchison2013, Lather2019, SimpkinsScience, EbbesenNMR, Murakoshi}.

Motivated by this progress, recent studies have proposed coupling a microwave cavity field to the anharmonic rotations of a single polar molecule with large electric dipole moments ($1-10\,$D) and transition frequencies in the $\mathrm{GHz}$ regime~\cite{Fan_2023,fanPulseareaTheoremPrecision2023,fanMaximizingOrientationThreestate2025}. Exploiting the nonlinear cavity physics encoded in this multilevel ladder requires strong light-matter coupling, a regime that is becoming increasingly accessible through rapid advances in molecular cavity QED and microwave resonator technology~\cite{Rabl2006,André2006NatPhys,Leghtas2015Science,Touzard2018PRX,Grimm2020Nature,Andr2006,FornDiaz2019,rubínosanz2026optimizingmagneticcouplinglumped,kanagin2025impuritiescryogenicsolidsnew}.

Most existing theoretical descriptions, however, simplify the rotational structure, reducing molecules to effective two-level systems~\cite{Fan_2023,fanPulseareaTheoremPrecision2023,fanMaximizingOrientationThreestate2025} and describing ensembles in terms of Dicke- or Tavis-Cummings-type collective variables~\cite{Rabl2006,André2006NatPhys,Andr2006}. Other directions focus on the control and spectroscopy of a single molecule in the strong-coupling regime~\cite{Albert2020Molecules,Bohn2017Science,RevModPhys.91.035005,Lemeshko_chapter,PhysRevLett.130.103202}. In both cases, the light-coupled Hilbert space is not constructed in a way that simultaneously preserves the multilevel rotational ladder and collective enhancement. 

\begin{figure}[t]
    \centering
    \includegraphics[width=\linewidth]{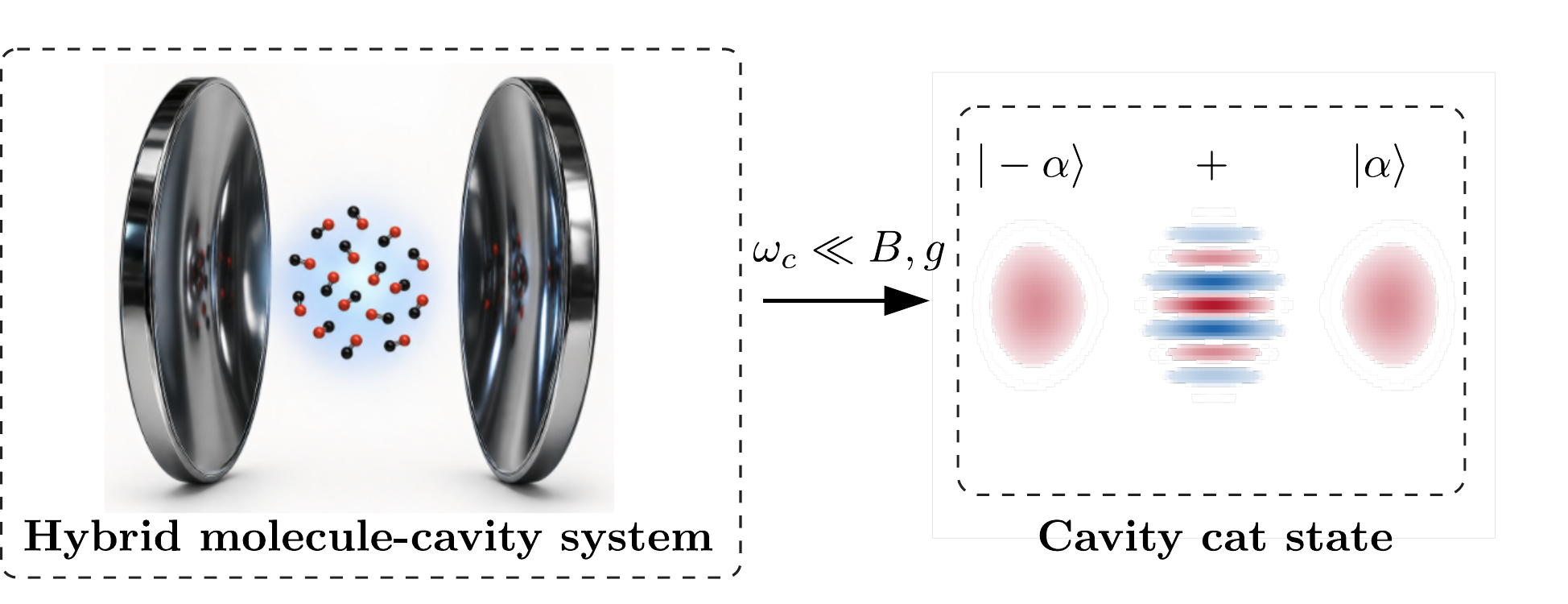}
    \caption{Schematic of the molecule-cavity system. A single cavity mode, described by the Hamiltonian $H_\mathrm{cav}=\omega_c \hat a^\dagger \hat a$, couples to rotational transitions of an ensemble of polar molecules with Hamiltonian $H_\mathrm{mol}=B\sum_i \hat{\mathbf L}_i^2$. In the bright collective manifold, multilevel virtual transitions along the rotational ladder generate effective two-photon and Kerr nonlinearities for the cavity field, hosting parity-protected cavity cat states.}
    \label{fig:fig1}
\end{figure}

In this Letter, we show that combining collective cavity coupling with the full rotational ladder gives rise to hybrid light-matter cat states, see Fig.~\ref{fig:fig1}. Our starting point is a first-principles treatment of the collective rotational structure. For homogeneous coupling to the cavity field and symmetric initial conditions, the cavity accesses a bright manifold generated by repeated action of the total polarization operator on the rotational vacuum. Constructing this manifold using a Krylov procedure~\cite{nandy2025quantum} yields a tridiagonal bright chain with $\sqrt{N}$-enhanced matrix elements while retaining the underlying rotor ladder. This provides a controlled reduction of the many-molecule problem to an effective collective bright rotor with enhanced coupling strength $g=\sqrt{N}g_0$, where $g_0$ denotes the single-molecule vacuum coupling. After introducing the rotational constant $B$ as the bare rotor energy scale, thresholds expressed in terms of $g/B$ translate directly into critical molecule numbers at fixed single-molecule coupling.

To the best of our knowledge, this work provides the first theoretical derivation of collective rotor-cavity coupling via an explicit Krylov bright-manifold construction~\cite{nandy2025quantum}, which we use to identify dispersive coupling to the collective rotational states as a route to nonclassical cavity cat states in the strong coupling regime. When the cavity is detuned from rotational transitions, real energy exchange and dissipation are suppressed, and the interaction proceeds through virtual transitions. The resulting state-dependent dispersive shifts~\cite{Boissonneault2009} enable quantum non-demolition spectroscopy with reduced backaction~\cite{Schuster2007}. We show that the anharmonicity of molecular rotation induces both an effective two-photon squeezing term and a Kerr nonlinearity for the cavity field. Their competition produces low-energy states with a cat-like phase-space structure, and it locks the parity of the photonic cat state to that of the collective rotational sector. 

We establish this picture in two complementary ways. Exact diagonalization of the collective bright-rotor Hamiltonian reveals a sharp crossover into a strongly correlated regime as well as cat-like Wigner functions in both the full cavity sector and the angular-momentum conditioned cavity state. We confirm this analysis by constructing a multi-level variational Ansatz and an effective cavity Hamiltonian from a fourth-order Schrieffer-Wolff expansion. Finally, rotor and cavity spectral functions connect the theoretical predictions to experimental observables. 

\textit{Model.} We consider $N$ identical bosonic polar molecules in a single-mode microwave cavity. Each molecule is a linear rigid rotor with rotational constant $B$ and permanent dipole moment $\mu$. We assume a linearly polarized cavity mode along the $\hat{\mathbf z}$ direction, work in units of $\hbar=1$, and use the long-wavelength dipole approximation. In the multipolar gauge, the Hamiltonian as derived in the Supplement~\cite{sup} reads
\begin{equation}
\hat H_N
=
B\sum_{i=1}^N \hat{\mathbf L}_i^2
+\omega_c \hat a^\dagger \hat a
+g_0(\hat a+\hat a^\dagger)\hat X_{\rm tot}
+\frac{g_0^2}{\omega_c}\hat X_{\rm tot}^2 ,
\label{eq:HN_model}
\end{equation}
where we introduced the total rotor polarization $\hat X_{\rm tot}$, bare cavity frequency $\omega_c$, coupling strength $g_0$ and $E_{\rm zpf}$ as the zero-point amplitude of the cavity field,
\begin{equation}
\hat X_{\rm tot}\equiv \sum_{i=1}^N \cos\hat\theta_i,\quad
g_0\equiv \mu E_{\rm zpf},\quad
E_{\rm zpf}=\sqrt{\frac{\omega_c}{2\varepsilon_0 V}} .
\label{eq:Xtot_g0}
\end{equation}
The parameters $\theta_i$, $\varepsilon_0$ and $V$ denote the angle between molecule $i$ and the cavity polarization axis, the vacuum permittivity, and the cavity mode volume, respectively. The term proportional to $\hat X_{\rm tot}^2$ is the dipole self-energy required for gauge invariance and ground-state stability in a single-mode description~\cite{Rokaj2018}.
\begin{figure}[t]
\includegraphics[width=\columnwidth]{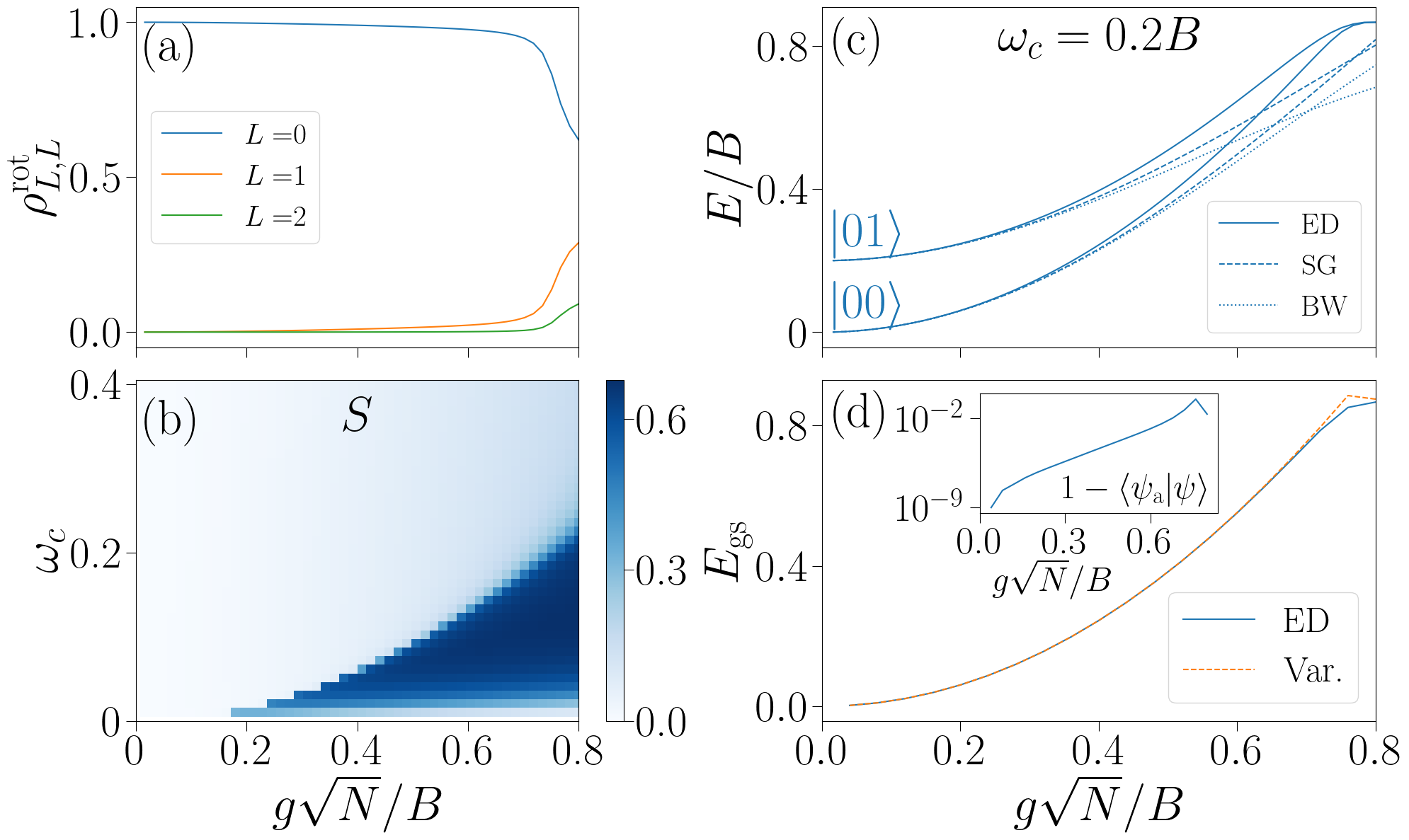}
    \vskip -0.3cm
    \caption{Exact diagonalization (ED) of an off-resonant cavity coupled to the collective bright rotational manifold as described by $\hat H_{\mathrm{coll}}$. (a) Diagonal elements of the reduced rotor density matrix, $\rho_{L,L}^{\rm rot}$, showing the redistribution of rotor populations with increasing coupling $g/B$. (b) Rotor-cavity entanglement entropy $S$ in the $\omega_c$-$g$ plane, showing the onset of strong rotor-photon correlations. (c) Lowest two energy levels compared with Brillouin-Wigner (BW) and Schr{\" o}dinger (SG) perturbation theory in the weak-coupling regime. (d) Ground-state energy compared with the variational energy obtained from the Ansatz $|\psi_{\rm a}\rangle$ in Eq.~\eqref{eq:variational_ansatz}. The inset shows the overlap between the variational Ansatz and the exact diagonalization ground state.}
    \label{fig:fig2}
\end{figure}
The cavity couples to the molecular ensemble only through $\hat X_{\rm tot}$, whereas the kinetic term $B\sum_i \hat{\mathbf L}_i^2$ does not reduce to a single collective rotor in general. For homogeneous coupling and symmetric initial conditions, however, the light predominantly accesses a bright symmetric manifold generated by repeated action of $\hat X_{\rm tot}$ on the rotational vacuum $|0\rangle\equiv|l{=}0\rangle^{\otimes N}$. We construct this manifold using a Krylov (Lanczos) procedure~\cite{nandy2025quantum}. Denoting the orthonormal Krylov states by~$|L\rangle_{\rm K}$, the operator $\hat X_{\rm tot}$ acts tridiagonally,
\begin{equation}
\hat X_{\rm tot}|L\rangle_{\rm K}
=
c_{L+1}|L{+}1\rangle_{\rm K}
+c_L|L{-}1\rangle_{\rm K},
\label{eq:Krylov_tridiag}
\end{equation}
with the lowest couplings being,
\begin{equation}
c_1=\sqrt{\frac{N}{3}},\qquad
c_2=\sqrt{\frac{10N-6}{15}},\qquad \ldots
\label{eq:beta12_main}
\end{equation}
so that $c_L\sim \sqrt{N}$. This construction identifies the full light-accessible bright manifold without truncating the rotational ladder to an effective two-level system. Projecting $\hat{H}_N$ onto the Krylov chain amounts to replacing
\begin{equation}
\hat X_{\rm tot}\ \mapsto\ 
\sum_{L\ge 0}c_{L+1}\big(|L\rangle_{\rm K}\langle L{+}1|+\text{h.c.}\big),
\label{eq:Xtot_bright_chain}
\end{equation}
which yields the bright-sector Hamiltonian
\begin{equation}
\begin{aligned}
\hat H_{\mathrm{coll}}
&=
\omega_c\,\hat a^\dagger \hat a
+\sum_{L\ge 0} E_L\,|L\rangle\langle L|\\
&+g_0(\hat a+\hat a^\dagger)\sum_{L\ge 0}c_{L+1}\Big(|L\rangle\langle L{+}1|+\mathrm{h.c.}\Big)\\
&+\frac{g_0^2}{\omega_c}\left[\sum_{L\ge 0}c_{L+1}\Big(|L\rangle\langle L{+}1|+\mathrm{h.c.}\Big)\right]^2 .
\label{eq:H_collective_rotor}
\end{aligned}
\end{equation}
The explicit construction and the role of dark multiplicity sectors are discussed in the Supplement~\cite{sup}. In the low-$L$ sector relevant below, we use $c_L\simeq \sqrt{N}\,c_L^{(0)}$ and $E_L\simeq B L(L+1)$. Physically, for spatially homogeneous coupling, the cavity addresses only the totally symmetric bright combination of molecular dipoles, so transition amplitudes from $N$ indistinguishable molecules add coherently and the matrix elements of $\hat X_{\rm tot}$ scale as $\sqrt{N}$. The remaining $N-1$ orthogonal combinations are dark to the cavity field and therefore do not contribute to the low-energy light-coupled dynamics. In the Krylov construction, this appears explicitly through the scaling $c_L\sim\sqrt{N}$ and justifies a collective coupling $g=\sqrt{N}g_0$. The same enhancement can be obtained by transforming to the collective polarization coordinate of the molecular rotors under the assumption that all molecules are aligned and rotate with similar angular velocities, cf.~\cite{buschTwoColdAtoms,rokajCavityInducedCollective2023} and the Supplement~\cite{sup}. 

Within the effective bright rotor Hamiltonian, thresholds stated in terms of $g/B$ therefore translate into critical molecule numbers $N\simeq (g/g_0)^2$ at fixed $g_0$. For Fabry-Pérot cavities, the mode volume is typically of the order of $V \sim 10^{-15}\,\mathrm{m}^3$~\cite{Rempe92,FornDiaz2019}. Polar molecules have dipole moments and rotational constants in the ranges $\mu \sim 1$--$10\,\mathrm{D}$ and $B / (2\pi) \sim 1$--$50\,\mathrm{GHz}$; for example, carbon monosulfide (CS) has $\mu \sim 2\,\mathrm{D}$ and $B \sim 24\,\mathrm{GHz}$~\cite{mocklerMicrowaveSpectrumCarbon1955}. In a dispersive cavity with $\omega_c / (2\pi) = 1\,\mathrm{GHz}$, these parameters correspond to a single-molecule coupling strength of $g_0 / (2\pi) \sim 0.1$--$1\,\mathrm{MHz}$. Reaching the strong-coupling regime, defined by $g/\omega_c \gtrsim 0.1$, therefore requires collective enhancement from ensembles of approximately~$N \sim 10^4 - 10^6$ molecules, similarly to polaritonic chemistry experiments~\cite{EbbesenRubioScholes, Hutchison2012, Thomas2016, EbbesenNMR, SimpkinsScience}. Alternative routes toward strong rotor-cavity coupling include reducing the mode volume, for example with plasmonic nanocavities~\cite{chikkaraddySinglemoleculeStrongCoupling2016,ojambatiQuantumElectrodynamicsRoom2019}, or employing parametric amplification methods~\cite{Qin2018}.

We diagonalize the bright Hamiltonian $\hat H_{\mathrm{coll}}$ from Eq.~\eqref{eq:H_collective_rotor} in the off-resonant regime $\omega_c\ll 2B$, where resonant exchange with the $0\!\leftrightarrow\!1$ angular momentum transition is suppressed and the dynamics is governed by virtual processes~\cite{Schuster2007}. At weak coupling, the ground state is perturbatively connected to $|L{=}0,n{=}0\rangle$. Increasing $g/B$ mixes higher rotor and photon sectors [Fig.~\ref{fig:fig2}(a)], and the rotor-cavity entanglement rises sharply beyond a critical cavity-frequency-dependent coupling strength~[Fig.~\ref{fig:fig2}(b)]. The same coupling scale marks the breakdown of second-order Schr{\" o}dinger and Brillouin-Wigner perturbation theory for the spectrum [Fig.~\ref{fig:fig2}(c)], indicating that the ground state is no longer a weakly dressed rotor-photon vacuum.

\begin{figure}
\includegraphics[width=\columnwidth]{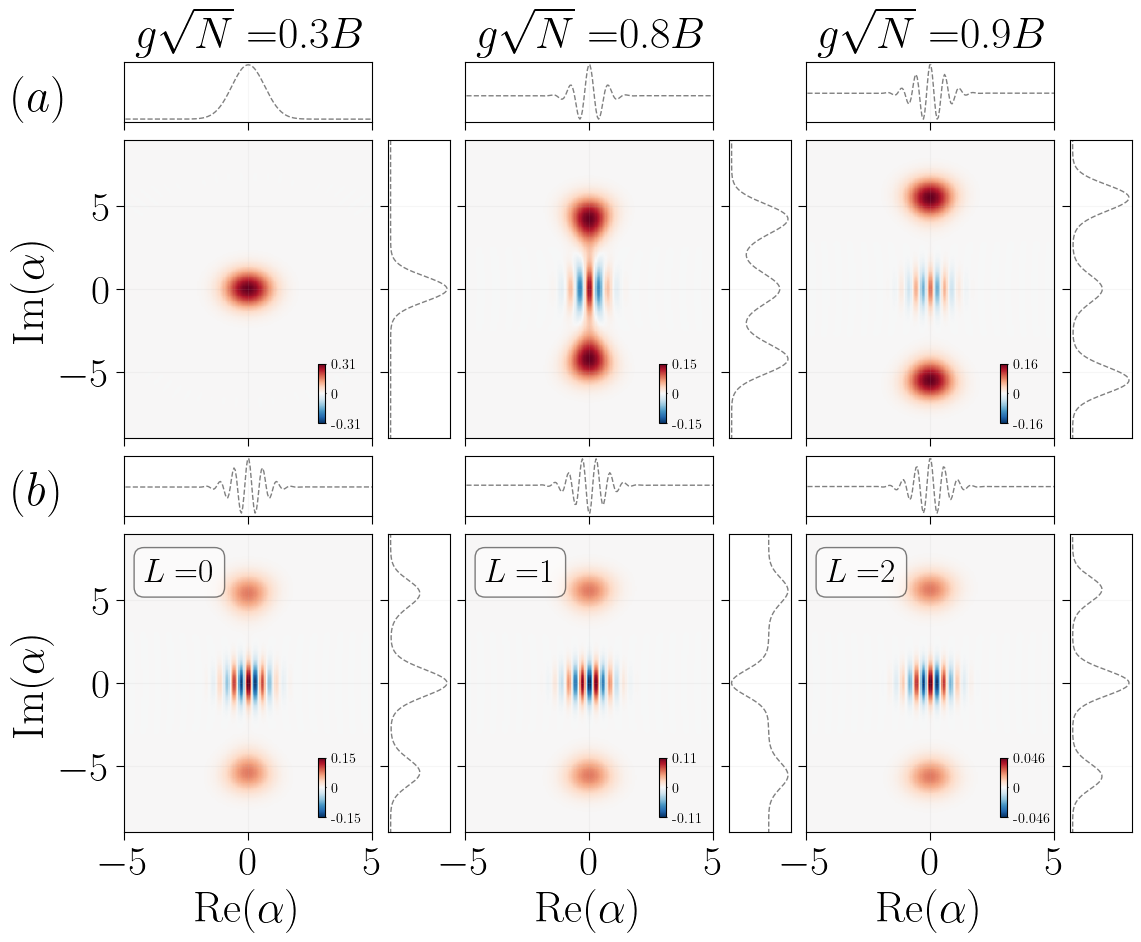}
    \vskip -0.3cm
    \caption{Wigner functions of the reduced cavity state in the ground-state manifold. (a) Wigner functions of $\rho^{\rm cav}$ for different coupling strengths at $\omega_c = 0.1B$. (b) Wigner functions of the angular-momentum-conditioned reduced cavity density matrix $\rho^{\rm cav}_L$ for $g = 0.8B$ and $\omega_c = 0.1B$. Even values of $L$ correspond to even Schr{\"o}dinger cat states, while odd values correspond to odd cat states.}
    \label{fig:fig3}
\end{figure}
The onset of the strongly correlated regime results in the emergence of cat state structure in the cavity sector. This is seen in the Wigner functions of the reduced cavity density matrix,
$\rho^{\rm cav} = \mathrm{Tr}_L \, |\psi\rangle \langle \psi|$,
obtained by tracing over all angular-momentum states [Fig.~\ref{fig:fig3}(a)]. With increasing coupling, the cavity state evolves from a Fock-like state into a Schr{\"o}dinger cat state. To resolve its parity structure, we examine cavity states conditioned on a fixed angular momentum, $\rho^{\rm cav}_L = |\psi_L\rangle \langle \psi_L|$. In the strongly correlated regime, these conditioned states become bimodal in phase space and exhibit the interference fringes characteristic of Schr{\" o}dinger cat states [Fig.~\ref{fig:fig3}(b)]. The cat parity is locked to the rotor parity, i.e. even $L$ correlates with an even cat and odd $L$ with an odd cat. This alternation follows from the dipole selection rule~$\Delta L=\pm 1$, which enforces the conservation of a joint rotor-photon parity in the bright manifold and organizes the effective Kerr and two-photon description derived below.

We capture this behavior with a variational Ansatz that assigns a parity-resolved cat to each angular-momentum sector,
\begin{equation}
|\psi_{\rm a}\rangle=\sum_L \beta_L\,|L\rangle\Big(|\alpha_L\rangle+(-1)^L |-\alpha_L\rangle\Big),
\label{eq:variational_ansatz}
\end{equation}
where $|\alpha_L\rangle$ are coherent states and $\beta_L$ are rotor weights. Minimizing the variational energy, $E_{\rm a}=\langle\psi_{\rm a}|\hat H|\psi_{\rm a}\rangle/\langle\psi_{\rm a}|\psi_{\rm a}\rangle$, with respect to $\{\alpha_L,\beta_L\}$ accurately reproduces the ground-state energy across the crossover and supports the interpretation of the cat regime in terms of parity-paired low-energy eigenstates [Fig.~\ref{fig:fig2}(d)].

Virtual multilevel excursions also leave clear dispersive signatures within linear response. In the following, we perform a Schrieffer-Wolff (SW) transformation to obtain an effective dispersive description with Stark shifts and, at higher order, photon pairing and Kerr nonlinearities. Because retaining the full rotor ladder is algebraically cumbersome, we introduce controlled truncations that isolate the relevant mechanisms.

\begin{figure}[h]
    \centering
    \includegraphics[width=\linewidth]{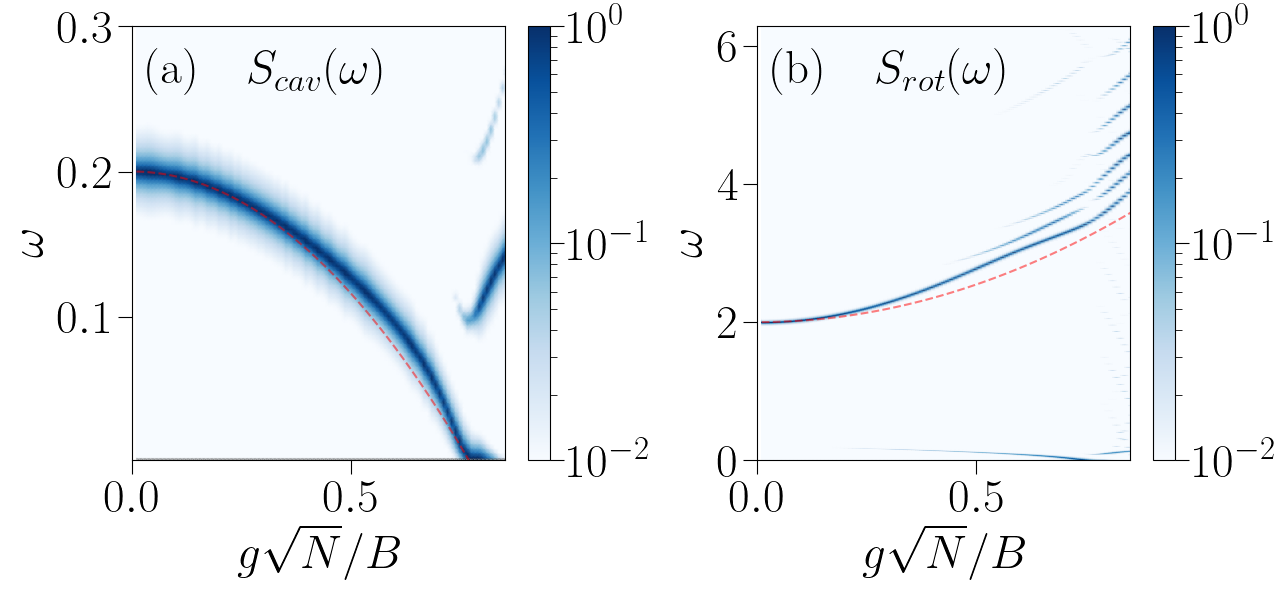}
    \vskip -0.3cm
    \caption{Spectral functions obtained from the time evolution of the respective ground state. (a) Cavity spectral function showing the dispersive peak shift and a polaritonic splitting at $g\sqrt{N}/B \sim 0.8$. (c) Rotor spectral function with corresponding Stark-shifted features. In the perturbative regime, dominant peak positions follow the SW dispersive description, while near the cat regime higher-order multilevel processes produce visible deviations. In both cases, analytic predictions are shown by red dashed lines, and the cavity is off-resonant at $\omega_c = 0.2B$.}
    \label{fig:fig4}
\end{figure}

In a two-level truncation to $L=0,1$, a second-order SW transformation gives
\begin{equation}
\begin{split}
    H_{2\times 2}^{\left(2\right)} =& \Big(\omega_c + 2 E_{\rm AC} \sigma_z\Big) \hat{a}^{\dagger} \hat{a} + \Big(\frac{\Delta}{2} + E_{\rm AC} \\
    &+ \frac{g^2 \delta\gamma}{2\omega_c}\Big) \sigma_z + E_{\rm AC} \sigma_z \left(\hat{a}^{\dagger} \hat{a}^{\dagger} + \hat{a} \hat{a}\right) + E_0 ,
\end{split}
\end{equation}
with $E_{\rm AC} = \frac{g^2\beta^2 \Delta}{\Delta^2 - \omega_c^2}$ and
$E_0 = \frac{g^2}{\omega_c} \gamma_{\rm av} + \frac{g^2\beta^2 \omega_c}{\Delta^2 - \omega_c^2}$. This effective Hamiltonian captures the dispersive cavity shift and the quadratic photon-pairing term, and it accounts for the Stark-shifted peak structure in the cavity and rotor spectral functions of the ground state~[Fig.~\ref{fig:fig4}].

However, the two-level truncation does not reproduce the nonlinear confinement in the cat regime, which arises from virtual transitions among higher rotational states. We therefore include a third level and expand the SW transformation to the fourth order, projecting onto the lowest bright manifold ($L=0$). The resulting effective cavity Hamiltonian is
\begin{equation}
\begin{aligned}
\label{eq:Heff_letter}
\hat H_{\rm eff}^{(L=0)}
=
E_0^{\rm (eff)}
+ A(\hat a^{\dagger 2}+\hat a^2)
+ \Omega\,\hat a^\dagger \hat a
+ \kappa(\hat a^{\dagger 4}+\hat a^4)\\
+ \mu(\hat a^{\dagger 3}\hat a+\hat a^\dagger \hat a^3)
+ \nu\,\hat a^{\dagger 2}\hat a^2
+\mathcal O(g^6),
\end{aligned}
\end{equation}
with coefficients given in the Supplement~\cite{sup}. The quadratic term $A(\hat a^{\dagger 2}+\hat a^2)$ provides an effective two-photon process, while $\nu\,\hat a^{\dagger 2}\hat a^2$ is the leading self-Kerr nonlinearity that confines the phase-space amplitude and occurs only for a system with more than two levels. Together, they form the minimal Kerr and two-photon structure needed to explain the parity-paired cat-like low-energy states in Fig.~\ref{fig:fig2}(d), while $\kappa$ and $\mu$ provide controlled higher-order corrections.

To test these analytical ac Stark shift predictions and further rotor-cavity hybridization, we compute the cavity and rotor spectral functions within linear response from the retarded Kubo response functions, which are given by
\begin{align}
\chi_{\rm cav}^{\rm R} (\omega) &= -i \int \mathrm{d} t\, e^{i\omega t} \, \theta (t) \langle [\hat a(t),\hat a^\dagger(0)] \rangle_0, \\
\chi_{\rm rot}^{\rm R} (\omega) &= -i \int \mathrm{d} t\, e^{i\omega t} \, \theta (t) \langle [\cos\theta(t),\cos\theta(0)] \rangle_0 .
\end{align}
The spectral functions are obtained from the respective imaginary parts, $S_{\rm cav/rot} (\omega) = - 2 \, \mathrm{Im} \chi_{\rm cav/rot}^{\rm R} $. The cavity spectrum probes photonic excitations, while the rotor spectrum captures rotational transitions mediated by the collective dipole operator $\cos\theta$ in the bright Krylov basis. Always considering the respective ground state, the dominant peaks in the numerically evaluated spectral functions follow the analytical predictions of the SW transformation at low couplings [Fig.~\ref{fig:fig4}]. As the coupling increases, the cavity mode softens and develops a polaritonic splitting at $g\sqrt{N}/B \sim 0.8$, whereas the rotor excitations are shifted to higher frequencies and acquire weak higher order transitions from $g\sqrt{N}/B \sim 0.5$. Both features provide experimentally measurable signatures of rotor-cavity entanglement. 

Although our analysis focuses on the closed-system low-energy manifold, the underlying mechanism is compatible with realistic cryogenic operations. Since rotational splittings are typically in the GHz range, thermal occupation of excited molecular states is already strongly suppressed at low temperatures. For representative species such as~CS, a temperature of $T\sim 1\,\mathrm{K}$ corresponds to a ground-state population close to unity. The cavity field is subject to a similar requirement, with thermal photons and the associated decoherence suppressed when $k_B T \ll \omega_c$. 

\textit{Discussion and outlook.} The mechanism identified here combines two ingredients that are natural to molecules, yet difficult to engineer with fabricated nonlinear elements: collective enhancement in the bright manifold and intrinsic multilevel anharmonicity of the rotational ladder. The experimental conditions to implement these ingredients are rapidly converging. On the molecular side, dense and long-lived polar molecule samples are now available, including quantum-degenerate gases enabled by microwave shielding and Bose-Einstein condensates of dipolar molecules~\cite{anderegg2021observation, PhysRevResearch.7.023164,softley2023cold}. The collective-state description assumes that rotational coherence survives over time scales longer than the inverse dispersive energy scales. That assumption is increasingly realistic in light of recent progress with ultracold molecules: rotational-state qubits in optical tweezers now exhibit coherence times from the $10^{-1}\,\mathrm{s}$ scale to the second scale, and both long-lived entanglement and multilevel coherent control of rotational manifolds have been demonstrated~\cite{PhysRevLett.127.123202,PhysRevLett.128.223201,gregory2024second}. These advances support treating the molecular sector as a coherent multilevel resource rather than as a strongly broadened bath. On the cavity side, superconducting microwave resonators routinely achieve millisecond photon lifetimes, with state-of-the-art devices extending into the tens-of-milliseconds regime~\cite{place2021new}. A realistic near-term implementation is therefore a hybrid architecture operated deep in the dispersive regime, in which the molecular ensemble primarily engineers the effective cavity nonlinearity and already modest cat sizes are sufficient to reveal the predicted parity structure.

This work opens several further directions. Elliptically polarized cavity modes or modes carrying orbital angular momentum can lift the degeneracy of the magnetic quantum number $M$ and generate bright rotor manifolds with additional internal structure~\cite{PhysRevResearch.6.033277}, providing another route for multi-component cat states. Coupling molecular ensembles to multimode resonators, or arranging them in arrays with controlled interactions, could further realize hybrid light-matter lattices in which strong photon nonlinearities are inherited directly from molecular rotation. More broadly, collective rotational-photonic cat eigenstates provide a molecular route to bosonic encodings and microwave molecule-photon interfaces that complements circuit QED while avoiding auxiliary qubits or engineered dissipation.

\textit{Acknowledgments.} We thank Ana M. Rey, Timur Tscherbul, Ceren Dağ, Dabayan Mitra, Mikhail Maslov, Christian Siegele, Georgios Koutentakis and Mateja Hrast for the stimulating discussions. V.R. acknowledges the financial support provided by the Villanova University Summer Grant program during the summer of 2026.

\bibliography{main} % Produces the bibliography via BibTeX.

\begin{widetext}
\section{Supplementary material}

\subsection{Derivation of the cavity coupling Hamiltonian for an ensemble of rotating dipoles}

In this section we derive the coupling of the cavity field to an ensemble of $N$ molecular dipoles from first principles. Each rotor $i$ is modeled as a pair of two charges $\pm q$ at fixed separation $\ell$, with center coordinate $\hat{\mathbf R}_i$ and orientation coordinate $\hat{\mathbf r}_i$~\cite{craig2012molecular},
\begin{equation}
\hat{\mathbf r}_{i\pm}=\hat{\mathbf R}_i \pm \frac{1}{2}\hat{\mathbf r}_i,
\qquad
|\hat{\mathbf r}_i|=\ell,
\qquad
\hat{\mathbf d}_i = q\,\hat{\mathbf r}_i \equiv \mu\,\hat{\mathbf n}_i,
\quad \mu=q\ell.
\label{eq:dipole_geometry}
\end{equation}
For clarity, the coordinate system is graphically illustrated in Fig.~\ref{fig:coordinate_system_description}. In the following, we eventually fix $\hat{\mathbf R}_i$ and keep only the rotational motion of $\hat{\mathbf n}_i$.
\begin{figure}[h]
    \centering
    \includegraphics[width=0.7\linewidth]{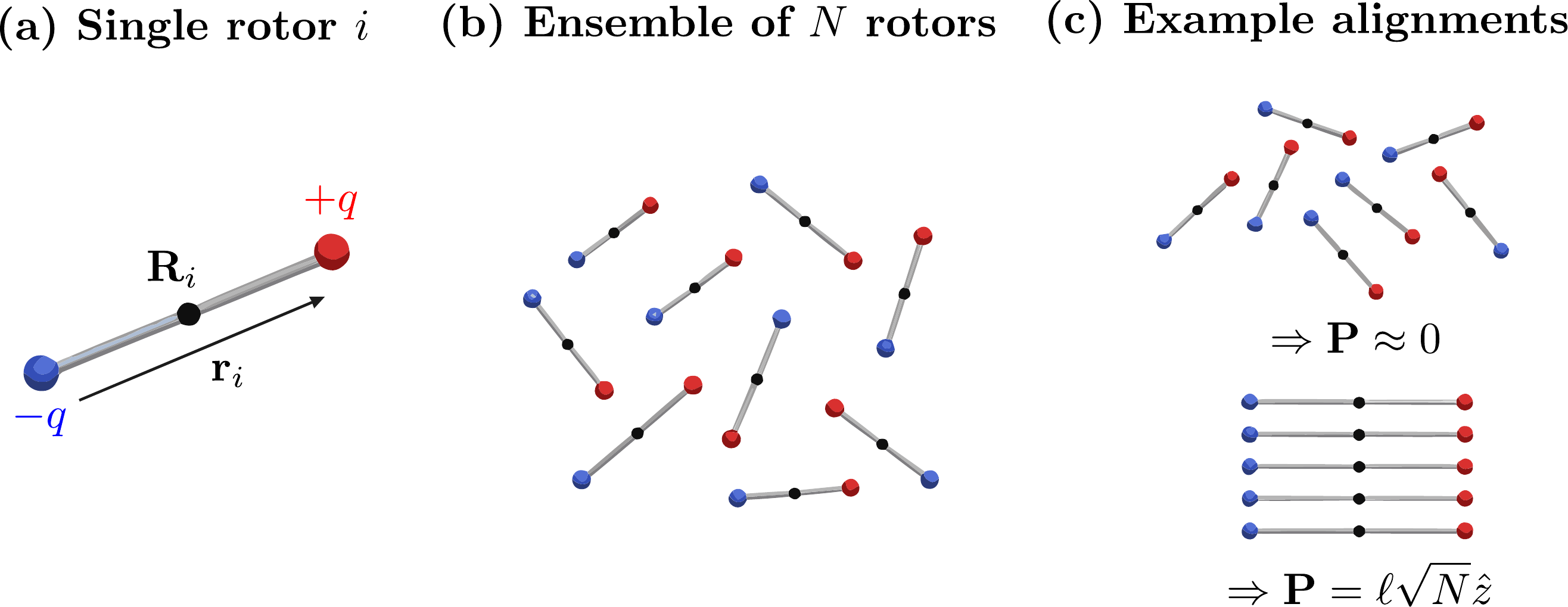}
    \vskip -0.3cm
    \caption{(a) Coordinate system for a single dipolar rotor: the vector $\mathbf{R}_i$ points towards the center of the molecule, while  $\mathbf{r}_i$ indicates its orientation and therefore the dipole direction, see Eq.~\eqref{eq:dipole_geometry}. (b) Ensemble of $N$ rotors, illustrating the collective polarization defined by the vector sum $\mathbf{P}=\sum_i \mathbf{r}_i / \sqrt{N}$, where the $\sqrt{N}$ factor is a normalization convention. (c) Representative configurations: for randomly oriented rotors, destructive interference leads to $\mathbf{P}\approx 0$, whereas for perfectly aligned dipoles the polarization scales with the number of molecules, $\mathbf{P}\propto \sqrt{N}$, reflecting a macroscopically ordered state.}
    \label{fig:coordinate_system_description}
\end{figure}

\noindent We briefly outline the standard transformation from the many-molecule  Hamiltonian in the Coulomb gauge to the Power-Zienau-Woolley (PZW) form  for this rotor model~\cite{craig2012molecular}. \\

\paragraph{Coulomb gauge Hamiltonian.} We start from the many-molecule Hamiltonian in the Coulomb gauge ($\nabla\cdot \mathbf{A} = 0$), 
\begin{equation}
\hat H_C
=
\sum_{i=1}^N\sum_{\sigma=\pm}
\left[
-\frac{\hbar^2}{2m}
\left(
\nabla_{i\sigma}-\frac{i\sigma q}{\hbar}\hat{\mathbf A}_\perp(\hat{\mathbf r}_{i\sigma})
\right)^2
\right]
+\hat V_{\mathrm{bind}}(\{|\hat{\mathbf r}_i|\})
+\hat H_{\mathrm f}
+\hat V_{\mathrm Coul}.
\label{eq:HC_rotor_charges}
\end{equation}
Here $\nabla_{i\sigma}\equiv \nabla_{\mathbf r_{i\sigma}}$ is the gradient operator with respect to particle coordinates, $m$ is the molecular mass, $\hat V_{\mathrm{bind}}(\{|\hat{\mathbf r}_i|\})$ is a binding potential and $\hat V_{\mathrm Coul}$ denotes the Coulomb interactions between charged particles. The cavity field Hamiltonian $\hat H_{\mathrm f}$ is described by,
\begin{equation}
\hat H_f
=
\frac{1}{2}
\int d^3 r\,
\left[
\varepsilon_0 \hat{\mathbf E}_{\perp}^2(\mathbf r)
+
\mu_0^{-1}\hat{\mathbf B}^2(\mathbf r)
\right],
\label{eq:cavity_bare_field_hamiltonian}
\end{equation}
where $\hat{\mathbf E}_{\perp}(\mathbf r)$ is the perpendicular electric field, $\hat{\mathbf B} (\mathbf r)$ is the magnetic field, $\varepsilon_0$ is the vacuum permittivity and $\mu_0$ is the vacuum permeability. \\

\noindent\paragraph{Long-wavelength limit.} In the electric-dipole (long-wavelength) limit, we assume that the electromagnetic field is constant over the extension of the rotors. This allows us to evaluate the transverse vector potential $\hat{\mathbf A}_\perp$ at the center position of the molecule,
\begin{equation}
\hat{\mathbf A}_\perp\!\left(\hat{\mathbf R}_i \pm \frac{1}{2}\hat{\mathbf r}_i\right)
\approx
\hat{\mathbf A}_\perp(\hat{\mathbf R}_i).
\label{eq:A_dipole_approx}
\end{equation}
At the same time, it is convenient to express derivatives with respect to the particle coordinates in terms of center and relative coordinates. Applying the chain rule, the gradient operators factorize into a gradient for the center position, and a gradient for the molecule orientaion, 
\begin{equation}
\nabla_{i\pm}=\frac{1}{2}\nabla_{\mathbf R_i}\pm \nabla_{\mathbf r_i}.
\label{eq:grad_R_r}
\end{equation}
Now we introduce the polarization density for rotor $i$ at position $\mathbf{x}$ following~\cite{craig2012molecular},
\begin{equation}
\hat{\mathbf P}_i(\mathbf x)
=
\sum_{\sigma=\pm}\sigma q\int_0^1 d\lambda\,
\left(\hat{\mathbf r}_{i\sigma}-\hat{\mathbf R}_i\right)
\delta\!\left(\mathbf x-\hat{\mathbf R}_i-\lambda(\hat{\mathbf r}_{i\sigma}-\hat{\mathbf R}_i)\right),
\label{eq:Pi_def}
\end{equation}
which describes the distribution of bound charge along the line connecting the two charges of the rotor. The integral over $\lambda$ parametrizes this line segment from the center position $\hat{\mathbf R}_i$ to the particle positions $\hat{\mathbf r}_{i\sigma}$, ensuring a continuous representation of the microscopic charge distribution. For point dipoles, this reduces to
\begin{equation}
\hat{\mathbf P}_i(\mathbf x)\approx \hat{\mathbf d}_i\,\delta(\mathbf x-\hat{\mathbf R}_i),
\qquad
\hat{\mathbf d}_i=q\hat{\mathbf r}_i.
\label{eq:Pi_dipole}
\end{equation}
\noindent\paragraph{PZW transformation.} The PZW transformation is realized by the unitary operator
\begin{equation}
\hat U=\exp\!\left[\frac{i}{\hbar}\int d^3x\,\hat{\mathbf P}(\mathbf x)\cdot\hat{\mathbf A}_\perp(\mathbf x)\right],
\qquad
\hat{\mathbf P}=\sum_i\hat{\mathbf P}_i.
\label{eq:U_PZW}
\end{equation}
In the dipole approximation, taking into account Eq.~\eqref{eq:A_dipole_approx} and Eq.~\eqref{eq:Pi_dipole}, this reduces to
\begin{equation}
\hat U\approx\exp\!\left[\frac{i}{\hbar}\sum_{i=1}^N \hat{\mathbf d}_i\cdot\hat{\mathbf A}_\perp(\hat{\mathbf R}_i)\right].
\label{eq:U_dipole}
\end{equation}
Applying the PZW identities from~\cite{craig2012molecular},
\begin{equation}
\hat U\,\hat{\mathbf A}_\perp(\mathbf x)\,\hat U^\dagger=\hat{\mathbf A}_\perp(\mathbf x),
\qquad
\hat U\,\hat{\boldsymbol\Pi}_\perp(\mathbf x)\,\hat U^\dagger
=
\hat{\boldsymbol\Pi}_\perp(\mathbf x)+\hat{\mathbf P}_\perp(\mathbf x).
\label{eq:Pi_shift}
\end{equation}
and fixed molecular centers,
\begin{equation}
\hat U\left(\nabla_{\mathbf r_i}-\frac{i q}{\hbar}\hat{\mathbf A}_\perp(\hat{\mathbf R}_i)\right)\hat U^\dagger
=
\nabla_{\mathbf r_i}.
\label{eq:grad_shift}
\end{equation}
Therefore, collecting all contributions in Eq.~\eqref{eq:HC_rotor_charges} that depend on the vector potential, the PZW transformation yields
\begin{equation}
\hat U \Big( \hat H_{\mathrm f} + \hat H_{\mathbf A} \Big) \hat U^\dagger
=
\hat H_{\mathrm f}
+ \hat H_{\mathrm{int}}
+ \frac{1}{2\varepsilon_0}\int d^3x\,\hat{\mathbf P}_\perp(\mathbf x)^2,
\label{eq:Hfield_to_PE_P2}
\end{equation}
where $\hat H_{\mathbf A}$ denotes all matter terms involving $\hat{\mathbf A}$. The dipole interaction term takes the form
\begin{equation}
\hat H_{\mathrm{int}} = -\int d^3x\,\hat{\mathbf E}_\perp(\mathbf x)\cdot\hat{\mathbf P}_\perp(\mathbf x)
\approx
-\sum_{i=1}^N \hat{\mathbf d}_i\cdot\hat{\mathbf E}_\perp(\hat{\mathbf R}_i).
\label{eq:Hint_dE}
\end{equation}
In particular, the coupling term to the transverse vector potential is transformed away and one has instead a coupling to the electric field $\mathbf{E}_\perp$. Contrastingly, the covariant kinetic terms transform trivially under $\hat U$. For molecule $i$, the sum of its two minimal-coupling kinetic contributions $\hat T_{i} = \sum_{\sigma=\pm} -\hbar^2 \nabla_{\mathbf{r}_i} / (2m)$ in Eq.~\eqref{eq:HC_rotor_charges} reduces to the bare Laplacian,
\begin{equation}
\hat U\,\hat T_{i}\,\hat U^\dagger
=
-\frac{\hbar^2}{m}\nabla_{\mathbf r_i}^2.
\label{eq:Ti_transformed}
\end{equation}
Here we assumed that the center positions of the molecules do not change and henceforth do not contribute to the kinetic energy, also cf.~Eq.~\eqref{eq:grad_R_r}. Collecting all contributions, we obtain
\begin{equation}
\hat H_{\mathrm{PZW}}
=
\hat{U} H_{C} U^{\dagger}
=
\sum_{i=1}^N
\left[
-\frac{\hbar^2}{m}\nabla_{\mathbf r_i}^2
+\hat V_{\mathrm{bind}}(|\hat{\mathbf r}_i|)
\right]
+\hat H_{\mathrm f}
-\sum_{i=1}^N \hat{\mathbf d}_i\cdot\hat{\mathbf E}_\perp(\hat{\mathbf R}_i)
+\int d^3x\,\frac{1}{2\varepsilon_0}\hat{\mathbf P}_\perp(\mathbf x)^2
+\hat H_{\mathrm{(multipoles)}}.
\label{eq:HPZW_final}
\end{equation}
Note that the remainder $\hat H_{\mathrm{(multipoles)}}$ contains contributions starting at electric quadrupole / magnetic dipole order. In the following two sections, Sec.~\ref{subsec:collective_coupling_to_total_polarization} and Sec.~\ref{subsec:collective_coupling_krylov}, we discuss two complementary routes to reduce the many-molecule PZW Hamiltonian in Eq.~\eqref{eq:HPZW_final} to an effective description where the quantized cavity field couples to a single collective degree of freedom with enhanced coupling strength. \\

\noindent\paragraph{Rigid rotor limit.} Let us now impose the rigid constraint $|\hat{\mathbf r}_i|=\ell$ and restrict to rotational degrees of freedom, so that the rotor kinetic energy becomes
\begin{equation}
{
\hat H_{\mathrm{rot}} = B\sum_{i=1}^N \hat{\mathbf J}_i^2,
\qquad B=\frac{\hbar^2}{2I}
}
\label{eq:Hrot_sumJi2}
\end{equation}
with moment of inertia $I$. The dipole moment $\hat{\mathbf d}_i=\mu\,\hat{\mathbf n}_i$ in the interaction term from Eq.~\eqref{eq:Hint_dE} has constant magnitude  and the self-energy is the projected $\int \hat{\mathbf P}_\perp^2/(2\varepsilon_0)$ term.
Let us now define the total angular momentum operator
\begin{equation}
\hat{\mathbf L}\equiv\sum_{i=1}^N \hat{\mathbf J}_i.
\end{equation}
Importantly, note that a ``center-of-mass rotation'' theory would require the kinetic term to depend only on $\hat{\mathbf L}^2$ (a single Laplacian on a single collective orientation variable). But already for $N=2$,
\begin{equation}
\hat{\mathbf J}_1^2+\hat{\mathbf J}_2^2=\frac{1}{2}\left(\hat{\mathbf L}^2+\hat{\mathbf K}^2\right),
\qquad
\hat{\mathbf K}\equiv \hat{\mathbf J}_1-\hat{\mathbf J}_2,
\label{eq:JK_identity}
\end{equation}
so the kinetic energy contains an independent relative operator $\hat{\mathbf K}^2$, which comes from the non-trivial angular momentum addition algebra.
For general $N$, we are free to choose an orthogonal matrix $O\in O(N)$ with first row $(1/\sqrt N,\dots,1/\sqrt N)$ and define collective/relative generators
\begin{equation}
\hat{\mathbf J}^{(\alpha)}\equiv \sum_{i=1}^N O_{\alpha i}\hat{\mathbf J}_i,
\qquad \alpha=0,1,\dots,N-1,
\end{equation}
so that $\hat{\mathbf J}^{(0)}=\hat{\mathbf L}/\sqrt N$ and $\hat{\mathbf J}^{(\alpha\ge 1)}$ are $(N-1)$ independent relative generators.
Then
\begin{equation}
{
\sum_{i=1}^N \hat{\mathbf J}_i^2
=
\sum_{\alpha=0}^{N-1} \left(\hat{\mathbf J}^{(\alpha)}\right)^2
=
\frac{1}{N}\hat{\mathbf L}^2
+
\sum_{\alpha=1}^{N-1}\left(\hat{\mathbf J}^{(\alpha)}\right)^2.
}
\label{eq:sumJi2_decomp}
\end{equation}
We conclude that unless physical constraints freeze all relative sectors $\alpha\ge 1$ (locking), $\hat H_{\mathrm{rot}}$ cannot be reduced to a function of $\hat{\mathbf L}^2$ only. Note that this is the precise analogue of ``relative coordinates do not decouple'' for free rotors. \\

\subsection{Collective coupling to the total polarization operator} \label{subsec:collective_coupling_to_total_polarization}        

In this section, we give an intuitive picture for the collective coupling to the total polarization operator. A detailed derivation of the collective coupling Hamiltonian is done in Sec.~\ref{subsec:collective_coupling_krylov}. \\

\paragraph{Collective polarization} Starting from the dipole-gauge light-matter interaction $\hat H_{\mathrm{int}}$ in Eq.~\eqref{eq:Hint_dE}, the molecular degrees of freedom enter through the rotor orientation vector $\hat{\mathbf r}_i = \ell\,\hat{\mathbf n}_i$ as defined in Eq.~\eqref{eq:dipole_geometry}. Since the total dipole moment is proportional to their sum, we introduce a normalized collective coordinate describing the total polarization,
\begin{equation}
    \tilde{\mathbf{P}} = \frac{1}{\sqrt{N}} \sum_{i=1}^N \hat{\mathbf{r}}_i
    = \frac{\ell}{\sqrt{N}} \sum_{i=1}^N \hat{\mathbf n}_i
    = \frac{\ell}{\sqrt{N}} \hat{\mathbf N}, \label{eq:normalized_coll_polarization}
\end{equation}
together with $N-1$ relative orientation coordinates
\begin{equation}
    \tilde{\mathbf{P}}_j = \frac{\mathbf r_1 - \mathbf r_j}{\sqrt{N}}
    = \frac{\ell}{\sqrt{N}}(\hat{\mathbf n}_1 - \hat{\mathbf n}_j),
    \qquad j>1.
\end{equation}
We note that the $1/\sqrt{N}$ factor normalizes the collective coordinate so that the associated collective moment of inertia in the kinetic energy scales like the one of a single rotor. Without the $1/\sqrt{N}$ factor, the collective moment of inertia analog would scale linearly with $N$, cf.~\cite{buschTwoColdAtoms,rokajCavityInducedCollective2023}

\begin{comment}
It comes from the permutation combinations of bosonic particles and corresponds to a quantum enhancement. In the Dicke model~\cite{FornDiaz2019}, for example, the normalized collective bright state for $N$ atoms is $|B\rangle = \sum_{j=1}^N |g_1,\dots,e_j,\dots,g_N\rangle / \sqrt{N}$ and the ground state is $|G\rangle =  |g_1,\dots,g_N\rangle$, with $g_i$ and $e_i$ denoting that atom $i$ is in the ground or excited state. Then, the matrix element of the total dipole operator acquires a $\sqrt{N}$ enhancement, $\langle G| \sum_{j=1}^N \mathbf{d}_j | B\rangle = \sqrt{N} \mathbf{d}_{e\to g}$ with Dicke dipole operators $\mathbf{d}_j$ and $\mathbf{d}_{e\to g}$. Therefore, we can put the $\sqrt{N}$ factor also in the definition of the total polarization, as written in Eq.~\eqref{eq:normalized_coll_polarization}. \\
\end{comment}

Going back to the physical interpretation of the polarization, the operator $\hat{\mathbf N} = \sum_i \hat{\mathbf n}_i$ measures the collective alignment of the rotor ensemble. For instance, if all rotors align along the $\hat z$ axis, $\hat{\mathbf N} = N \hat z$, whereas for antiparallel or random orientations $\hat{\mathbf N} \approx 0$, implying a vanishing net dipole moment and consequently no coupling to the cavity field, see Fig.~\ref{fig:coordinate_system_description}. Importantly, the magnitude of $\hat{\mathbf N}$ is not conserved, and the problem is  not anymore spherically symmetric in this collective coordinate. Moreover, for rigid rotors, $|\hat{\mathbf n}_i|=1$, this coordinate transformation is non-orthogonal. Because of the multi-level structure of molecular rotations, the many-body rotor states are more complicated than in the Dicke model. In the following, by working in the low-energy limit, we give an intuitive construction of the many-body rotor states that are connected by repeated action of the collective polarization operator. For a detailed derivation that includes the matrix elements of the polarization and kinetic energy operators in the many-body basis, we refer to the Krylov construction described in Sec.~\ref{subsec:collective_coupling_krylov}. \\

\paragraph{Low-energy limit} In the low-energy sector, all rotors are in their angular momentum ground state with $l=0$, which we call the rotational vacuum state. Starting from this state, we construct the many-body rotor Hilbert space by introducing an occupation-number representation under the assumption of bosonic molecules,
\begin{equation}
    | \{n_l\} \rangle \equiv |n_0, n_1, n_2, \dots \rangle,
\end{equation}
where $n_l$ denotes the number of molecules occupying the single-rotor angular momentum sector $l$ (with $m=0$ throughout). The constraint $\sum_l n_l = N$ is implied. In this notation, the lowest-energy states can be organized in ascending order of their total rotational energy,
\begin{equation}
\begin{split}
    |\psi_{g}\rangle &= |N,0,0,0,\dots\rangle, \qquad E_g = 0, \\
    |\psi_{1e}\rangle &= |N-1,1,0,0,\dots\rangle, \qquad E_{1e} = 2B, \\
    |\psi_{2e}\rangle &= |N-2,2,0,0,\dots\rangle, \qquad E_{2e} = 4B, \\
    |\psi_{3e}\rangle &\in \mathrm{span}\{ |N-1,0,1,0,\dots\rangle,|N-3,3,0,0,\dots\rangle\}, \qquad E_{3e} = 6B, \\
    |\psi_{4e}\rangle &\in \mathrm{span}\{ |N-1,0,0,1,\dots\rangle,\; |N-2,0,2,0,\dots\rangle,\; |N-6,6,0,0,\dots\rangle,\dots \},\quad E_{4e} = 12B,
\end{split}
\end{equation}
and so on. At a given excitation energy, the degeneracy arises from the number of ways to distribute angular momentum quanta among the rotors under the constraint $\sum_l n_l = N$. We now evaluate matrix elements of the collective orientation operator $\hat{\mathbf N}$ within this basis. Writing the components of $\hat{\mathbf n}_i$ as
$
\hat n_{i,x} = \sin\theta_i \cos\phi_i,
\hat n_{i,y} = \sin\theta_i \sin\phi_i,
\hat n_{i,z} = \cos\theta_i,
$
and using the spherical harmonic matrix elements,
\begin{equation}
    \langle \ell_1 | \hat n_{x,y} | \ell_2 \rangle = 0, \qquad
    \langle \ell_1 | \hat n_z | \ell_2 \rangle \neq 0 \;\; \text{only for } \ell_1=\ell_2\pm 1,
\end{equation}
we find that only the $z$-component survives due to the restriction to $m=0$. Consequently, the operator $\hat{\mathbf N}$ only connects states that differ by angular momentum $\Delta l = \pm 1$. Starting from the rotational vacuum, this implies that $\hat{\mathbf N}$ only generates transitions within the subset
\begin{equation}
    |N,0,0,\dots\rangle \;\leftrightarrow\; |N-1,1,0,\dots\rangle \;\leftrightarrow\; |N-2, 2, 0, 0, \dots\rangle \;\leftrightarrow\; \dots,
\end{equation}
which we identify as the \emph{bright manifold} of collective rotor states, in direct analogy to bright states in the Dicke model. All remaining states are dark with respect to the polarization operator $\hat{\mathbf N}$. Within this bright manifold, the collective orientation operator is a tridiagonal matrix similar in structure to $\cos\theta$. For an explicit construction, we refer to Sec.~\ref{subsec:collective_coupling_krylov}. \\

\begin{comment}
In the opposite limit of high excitation energy, there can be a scenario where all rotors are aligned in the same angular direction and move coherently. Then, $\hat{\mathbf N}$ approaches a quasi-classical vector with approximately fixed length and many angular momentum sectors contribute. In this work, we focus on the collective coupling of ultracold rotors which are in their rotational ground state. Therefore, this high energy limit is beyond the scope of our study. 

We conclude that the collective polarization operator $\tilde{\mathbf P}$ differs qualitatively from the usual collective center-of-mass coordinate encountered in other polaritonic systems~\cite{rokajCavityInducedCollective2023}. In particular, its magnitude is not conserved and therefore, it does not allow a simple separation into collective and relative kinetic energies. Nonetheless, within the low-energy sector, it generates a well-defined bright manifold. In the following, we formalize this intuitive picture within the Krylov basis framework.
\end{comment}

\subsection{Collective coupling from the Krylov basis} \label{subsec:collective_coupling_krylov}

This section formalizes the construction of the effective collective Hamiltonian from Eq.~\eqref{eq:H_collective_rotor} as sketched in Sec.~\ref{subsec:collective_coupling_to_total_polarization} by constructing a Krylov basis. \\

\paragraph{Second quantization of the cavity mode} When the cavity mode function is uniform on the scale of the the rotor ensemble, all rotors see the same field amplitude. Then the center positions $\hat{\mathbf R}_i$ drop out of $\hat H_{\mathrm{int}}$ and all coupling terms are equal. As mentioned above, the orientational relative sectors remain in $\hat H_{\mathrm{rot}}$ via Eq.~\eqref{eq:sumJi2_decomp}. This is the central structural difference compared to coupling to a collection of charges.
We now consider a single cavity mode of frequency $\omega_c$ with
\begin{equation}
\hat{\mathbf E}_\perp(\mathbf r) = \mathbf e\,E_{\mathrm{zpf}}\,f(\mathbf r)\,(\hat a+\hat a^\dagger),
\end{equation}
and assume $f(\hat{\mathbf R}_i)=1$ for all $i$, corresponding to identical values of
the cavity mode function at all rotor centers. The field zero-point amplitude is $E_{\mathrm{zpf}} = \sqrt{\omega_c / (2\varepsilon_0 V)}$, with $V$ being the effective mode volume and $\varepsilon_0$ being the vacuum permittivity. Choosing the $\mathbf e=\hat z$ axis for the cavity polarization, and working in the $M=0$ sector for each rotor, we have $\hat{\mathbf n}_i\cdot \hat z = \cos\hat\theta_i$. Then the interaction term from Eq.~\eqref{eq:Hint_dE} becomes
\begin{equation}
{
\hat H_{\mathrm{int}} = -\hbar g\,(\hat a+\hat a^\dagger)\,\hat X_{\mathrm{tot}},
\qquad
\hat X_{\mathrm{tot}}\equiv \sum_{i=1}^N \cos\hat\theta_i,
\qquad
g\equiv \frac{\mu E_{\mathrm{zpf}}}{\hbar}.
}
\label{eq:Hint_Xtot}
\end{equation}
The rotor Hamiltonian is
\begin{equation}
{
\hat H_{\mathrm{rot}} = B\sum_{i=1}^N \hat{\mathbf J}_i^2.
}
\end{equation}
Optionally (for large couplings), we also can include the (single-mode projected) self-energy term from Eq.~\eqref{eq:HPZW_final},
\begin{equation}
\hat H_{\mathrm{self}} = \hbar\frac{g^2}{\omega_c}\,\hat X_{\mathrm{tot}}^2.
\end{equation}

\paragraph{Many-body rotor states in occupation number representation} For a single rotor restricted to the $M=0$ sector, we define the shorthand $|J\rangle\equiv|J,M=0\rangle$. In this basis, the polarization operator of every rotor is tridiagonal,
\begin{equation}
\cos\hat\theta
=
\sum_{J\ge 0} c_J\left(|J{+}1\rangle\langle J|+|J\rangle\langle J{+}1|\right),
\qquad
c_J=\frac{J+1}{\sqrt{(2J+1)(2J+3)}}.
\label{eq:cos_tridiag_again}
\end{equation}
In particular, the first two coefficients are $c_0=1/\sqrt3$ and $c_1=2/\sqrt{15}$. We now expand this single rotor basis to the permutation-symmetric $N$-rotor subspace, which is the relevant matter sector for identical light-matter couplings and symmetric initial conditions. This subspace is represented by bosonic occupation operators $\{\hat b_J\}_{J\ge 0}$, where \(\hat b_J^\dagger\) creates one rotor in the angular-momentum state \(|J\rangle\). The total number of rotors is fixed and consequently $\sum_{J\ge 0}\hat b_J^\dagger \hat b_J = N$. The corresponding symmetric occupation basis is
$|n_0,n_1,n_2,\dots\rangle$ with $\sum_{J\ge 0}n_J=N$, which includes the symmetric angular momentum vacuum,
\begin{equation}
|0\rangle_{\mathrm K}\equiv |N,0,0,\dots\rangle \equiv |J{=}0\rangle^{\otimes N}.
\label{eq:rotational_vacuum}
\end{equation}
The subscript indicates that we will subsequently construct the Krylov states from $|0\rangle_{\mathrm K}$. In this basis representation, any one-body single-rotor operator
$\hat O=\sum_{J,J'} O_{J'J}|J'\rangle\langle J|$ can be expressed with creation and annihilation operators, i.e.~$\hat O = \sum_{J,J'} O_{J'J}\hat b_{J'}^\dagger \hat b_J$. Applying it to the polarization operator $\cos\hat\theta$ from Eq.~\eqref{eq:cos_tridiag_again}, gives the exact operator identity on the symmetric subspace,
\begin{equation}
{
\hat X_{\mathrm{tot}} =
\sum_{J\ge 0} c_J\left(\hat b_{J+1}^\dagger \hat b_J + \hat b_J^\dagger \hat b_{J+1}\right).
}
\label{eq:Xtot_boson_hat}
\end{equation}
Similarly, the rotor Hamiltonian remains diagonal in the occupation basis,
\begin{equation}
{
\hat H_{\mathrm{rot}}=\sum_{J\ge 0} E_J\,\hat b_J^\dagger \hat b_J,
\qquad
E_J=BJ(J+1).
}
\label{eq:Hrot_boson_hat}
\end{equation}

\paragraph{Krylov construction} The Krylov construction systematically identifies the bright collective manifold by repeated action of the interaction operator. Starting from a reference state, here the fully symmetric rotational vacuum state, successive applications of $\hat X_{\mathrm{tot}}$ generate the states that are directly connected by the cavity field. In this basis, the projected interaction takes a tridiagonal Lanczos form, which makes the collective enhancement and the structure of the bright ladder explicit. \\

The first Krylov states are obtained by applying $\hat X_{\mathrm{tot}}$ to the angular momentum vacuum state as defined in Eq.~\eqref{eq:rotational_vacuum} and orthonormalizing at each step,
\begin{equation}
|1\rangle_{\mathrm K} \equiv \frac{1}{\beta_1}\hat X_{\mathrm{tot}}|0\rangle_{\mathrm K},
\qquad
|2\rangle_{\mathrm K} \equiv \frac{1}{\beta_2}\left(\hat X_{\mathrm{tot}}|1\rangle_{\mathrm K}-\beta_1|0\rangle_{\mathrm K}\right),
\end{equation}
where the Lanczos coefficients $\beta_1,\beta_2>0$ are fixed by normalization. Let us compute them explicitly as functions of $N$. From the angular momentum sum in Eq.~\eqref{eq:Xtot_boson_hat}, only the angular momentum $J=0$ term acts nontrivially on the rotational vacuum $|N,0,0,\dots\rangle$, 
\begin{align}
\hat X_{\mathrm{tot}}|N,0,0,\dots\rangle
&=
c_0\,\hat b_1^\dagger \hat b_0|N,0,0,\dots\rangle
=
c_0\,\sqrt{N}\,|N{-}1,1,0,\dots\rangle.
\end{align}
Therefore, the first Krylov state is
\begin{equation}
{
|1\rangle_{\mathrm K}=|N{-}1,1,0,\dots\rangle,
\qquad
\beta_1=c_0\sqrt{N}=\sqrt{\frac{N}{3}}.
}
\label{eq:beta1_hat}
\end{equation}
This is the first sign of collective enhancement, as the matrix element from the symmetric vacuum to the first bright state scales as $\sqrt{N}$. Next, applying $\hat X_{\mathrm{tot}}$ to $|1\rangle_{\mathrm K}$ gives
\begin{align}
\hat X_{\mathrm{tot}}|N{-}1,1,0,\dots\rangle
&=
c_0(\hat b_1^\dagger \hat b_0+\hat b_0^\dagger\hat b_1)|N{-}1,1,0,\dots\rangle
+
c_1(\hat b_2^\dagger\hat b_1+\hat b_1^\dagger\hat b_2)|N{-}1,1,0,\dots\rangle\nonumber\\
&=
c_0\sqrt{2(N-1)}\,|N{-}2,2,0,\dots\rangle
+
c_0\sqrt{N}\,|N,0,0,\dots\rangle
+
c_1\,|N{-}1,0,1,0,\dots\rangle.
\end{align}
Subtracting the back component $\beta_1|0\rangle_{\mathrm K}=c_0\sqrt N|N,0,0,\dots\rangle$, the new orthogonal vector is
\begin{equation}
|\chi_2\rangle \equiv
c_0\sqrt{2(N-1)}\,|N{-}2,2,0,\dots\rangle
+
c_1\,|N{-}1,0,1,0,\dots\rangle.
\end{equation}
Its norm determines the second coefficient,
\begin{equation}
\beta_2^2 = \langle \chi_2|\chi_2\rangle
=
2(N-1)c_0^2+c_1^2
=
\frac{2(N-1)}{3}+\frac{4}{15}
=
\frac{10N-6}{15}
\end{equation}
which leads to
\begin{equation}
{
|2\rangle_{\mathrm K}
=
\frac{
c_0\sqrt{2(N-1)}\,|N{-}2,2,0,\dots\rangle
+
c_1\,|N{-}1,0,1,0,\dots\rangle
}{\beta_2},
\qquad
\beta_2=\sqrt{\frac{10N-6}{15}}.
}
\label{eq:beta2_hat}
\end{equation}
We can continue this construction to higher states Krylov states in a similar way. Within the first three Krylov states which define the manifold $\mathcal K^{(2)} = \mathrm{span} \lbrace|0\rangle_{\mathrm K}, |1\rangle_{\mathrm K},|2\rangle_{\mathrm K}\rbrace$, we define the projector
\begin{equation}
\hat P^{(2)}_{\mathrm K}\equiv \sum_{L=0}^2 |L\rangle_{\mathrm K}\,{}_{\mathrm K}\langle L|.
\end{equation}
By construction, the projected interaction is tridiagonal, 
\begin{equation}
{
\hat P^{(2)}_{\mathrm K}\,\hat X_{\mathrm{tot}}\,\hat P^{(2)}_{\mathrm K}
=
\beta_1\left(|0\rangle\langle 1|+|1\rangle\langle 0|\right)
+
\beta_2\left(|1\rangle\langle 2|+|2\rangle\langle 1|\right),
}
\label{eq:Xproj_hat}
\end{equation}
with coefficients $\beta_1$ and $\beta_2$ as given in Eq.~\eqref{eq:beta1_hat} and Eq.~\eqref{eq:beta2_hat}. Hence, the Krylov basis isolates the bright states generated by the collective dipole operator and expresses the interaction as a nearest-neighbor hopping problem along the bright ladder. \\

\paragraph{Rotor Hamiltonian in Krylov basis} A subtle point is that Krylov states such as $|2\rangle_{\mathrm K}$ in Eq.~\eqref{eq:beta2_hat} can contain angular-momentum configurations with different rotational energies. Therefore, the bright Krylov ladder is not identical to a single rigid collective rotor in general and the collective rotor problem is not analogous to the collective coupling to the charge center-of-mass. We compute $\hat H_{\mathrm{rot}}|2\rangle_{\mathrm K}$ explicitly using Eq.~\eqref{eq:Hrot_boson_hat}. From
\begin{equation}
\hat H_{\mathrm{rot}}|N{-}2,2,0,\dots\rangle = (2E_1)|N{-}2,2,0,\dots\rangle = 4B\,|N{-}2,2,0,\dots\rangle,
\end{equation}
and
\begin{equation}
\hat H_{\mathrm{rot}}|N{-}1,0,1,0,\dots\rangle = (E_2)|N{-}1,0,1,0,\dots\rangle = 6B\,|N{-}1,0,1,0,\dots\rangle,
\end{equation}
we obtain
\begin{equation}
\hat H_{\mathrm{rot}}|2\rangle_{\mathrm K}
=
\frac{1}{\beta_2}\left(4B\,\alpha_1\,|N{-}2,2,0,\dots\rangle + 6B\,\alpha_2\,|N{-}1,0,1,0,\dots\rangle\right),
\quad
\alpha_1\equiv c_0\sqrt{2(N-1)},\ \alpha_2\equiv c_1.
\label{eq:Hrot_on2_leak}
\end{equation}
Since the two components have different rotational energies, $4B\neq 6B$, the vector $\hat H_{\mathrm{rot}}|2\rangle_{\mathrm K}$ is not proportional to $|2\rangle_{\mathrm K}$, as $\hat H_{\mathrm{rot}}$ generates a component orthogonal to the one-dimensional bright state inside the two-excitation manifold. 
%as it has a component orthogonal to $\mathcal K^{(2)}$ within the fixed-$L=2$ multiplicity space. Equivalently,
The projected bright energy is
\begin{align}
E_2^{\mathrm{(br)}} &\equiv {}_{\mathrm K}\langle 2|\hat H_{\mathrm{rot}}|2\rangle_{\mathrm K}
=
\frac{4B|\alpha_1|^2+6B|\alpha_2|^2}{|\alpha_1|^2+|\alpha_2|^2}
=
4B\,\frac{5N-2}{5N-3},
\label{eq:E2bright_final}
\end{align}
and the first bright anharmonicity is
\begin{equation}
{
\alpha_N \equiv E_2^{\mathrm{(br)}}-2E_1
=
\frac{4B}{5N-3}\sim \frac{0.8\,B}{N}.
}
\label{eq:alphaN_final}
\end{equation}
This indicates that the bright Krylov ladder becomes harmonic in the large-$N$ limit, while retaining a finite-$N$ anharmonicity set by the microscopic rotor spectrum. \\

\paragraph{Full Hamiltonian in Krylov basis} We now combine the projected rotor kinetic energy and rotor-cavity interaction Hamiltonians. Starting from the single-mode Hamiltonian, temporarily neglecting the self-energy term,
\begin{equation}
\hat H = \hbar\omega_c \hat a^\dagger\hat a + \hat H_{\mathrm{rot}} - \hbar g(\hat a+\hat a^\dagger)\hat X_{\mathrm{tot}}
\end{equation}
we project onto the truncated Krylov space $\mathcal K^{(2)}$, 
\begin{equation}
{
\hat H_{\mathrm{br}}^{(2)} \equiv \hat P_{\mathrm K}^{(2)}\,\hat H\,\hat P_{\mathrm K}^{(2)}
=
\hbar\omega_c \hat a^\dagger\hat a
+
E_1\,|1\rangle\langle 1|
+
E_2^{\mathrm{(br)}}\,|2\rangle\langle 2|
-
\hbar g(\hat a+\hat a^\dagger)
\Big[
\beta_1(|0\rangle\langle 1|+\mathrm{h.c.})
+
\beta_2(|1\rangle\langle 2|+\mathrm{h.c.})
\Big],
}
\label{eq:Hbright_final}
\end{equation}
with variables
\begin{equation}
E_1=2B,\qquad E_2^{\mathrm{(br)}}=4B\frac{5N-2}{5N-3},\qquad
\beta_1=\sqrt{\frac{N}{3}},\qquad \beta_2=\sqrt{\frac{10N-6}{15}}.
\end{equation}
From $\hat H_{\mathrm{br}}^{(2)}$ in Eq.~\eqref{eq:Hbright_final}, the effective bright couplings are
\begin{equation}
g_{01}^{\mathrm{eff}} = g\,\beta_1 = g\sqrt{\frac{N}{3}},
\qquad
g_{12}^{\mathrm{eff}} = g\,\beta_2 \approx g\sqrt{\frac{2N}{3}}\quad (N\gg 1),
\end{equation}
These enhanced matrix elements are the rotor analogue of the charge center-of-mass coupling to a cavity mode. In both cases, the cavity couples to a collective coordinate, and the corresponding bright-state matrix elements scale as $\sqrt{N}$. \\

We illustrate the difference between the energy levels for the single and many-rotor case in Fig.~\ref{fig:diagonalization_of_bright_rotor}. For an ensemble with $N=10^8$ rotors, even a very small single-rotor coupling, of order $g\sim 10^{-5}B$, can generate a collectively enhanced interaction. Such values are compatible with Fabry-Perot cavities with typical mode volumes of the order $10^{3}\,\mu\mathrm{m}^3$. At increased coupling strengths, the reduced cavity state has a Wigner function profile in phase space, as analyzed in depth in the main text. The small population of the third level, $|L=2\rangle$, indicates that the bright Hamiltonian $\hat H_{\mathrm{br}}^{(2)}$ is a controlled approximation. Therefore, the single rotor results reported in the main text transfer directly to the many rotor case.

\begin{figure}[h]
    \centering
    \includegraphics[width=0.4\linewidth]{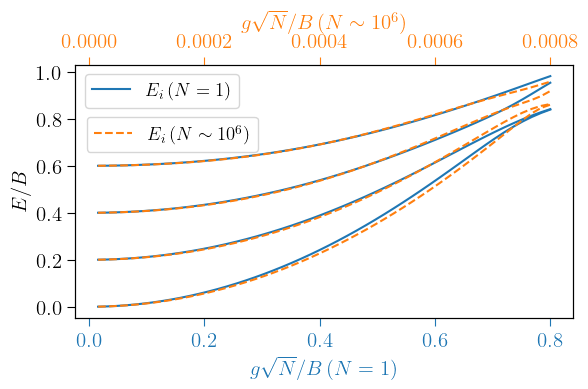}
    \vskip -0.3cm
    \caption{Energy levels of the bright Hamiltonian from Eq.~\eqref{eq:Hbright_final} with the rotor self-energy in the off-resonant regime, $\omega_c \ll B$, for $\omega_c = 0.2B$. Blue lines show the eigenenergies in the single rotor limit ($N=1$), while the dashed orange lines show the eigenergies for $N=10^8$ rotors. Note that the coupling strengths on the $x$ axis are rescaled accordingly.}
    \label{fig:diagonalization_of_bright_rotor}
\end{figure}

\paragraph{Leakage from the bright Krylov subspace} Unlike for the light-matter coupling to charged particles, where the center-of-mass and relative coordinates separate exactly and the collective subspace is invariant under the full Hamiltonian, the rotor kinetic energy $\hat H_{\mathrm{rot}}$ does not fully preserve the bright Krylov subspace. This can already be seen from the vector $\hat H_{\mathrm{rot}}|2\rangle_{\mathrm K}$, which is not proportional to $|2\rangle_{\mathrm K}$ and has orthogonal components to this bright state, see Eq.~\eqref{eq:Hrot_on2_leak}. Hence, the bright Hamiltonian $\hat H_{\mathrm{br}}^{(2)}$ in Eq.~\eqref{eq:Hbright_final} is a controlled effective model only in regimes where the cavity-driven dynamics remains dominantly within the space and leakage is perturbatively small, for example in the dispersive Schrieffer-Wolff regime where $|2\rangle$ is only virtually occupied. To make this leakage explicit, we construct a dark state orthogonal to $|2\rangle_{\mathrm K}$ within the same two-excitation occupation subspace. Let us introduce the shorthand notations $|\phi_a\rangle\equiv |N{-}2,2,0,\dots\rangle$ and $|\phi_b\rangle\equiv |N{-}1,0,1,0,\dots\rangle$, and recall the second Krylov state from Eq.~\eqref{eq:beta2_hat},
\begin{equation}
|2\rangle_{\mathrm K}=\frac{\alpha_1|\phi_a\rangle+\alpha_2|\phi_b\rangle}{\beta_2},\quad \alpha_1=c_0\sqrt{2(N-1)},\ \alpha_2=c_1,\ \beta_2=\sqrt{|\alpha_1|^2+|\alpha_2|^2}.
\end{equation}
An orthonormal dark state companion is then
\begin{equation}
|2\rangle_{\mathrm D} \equiv \frac{\alpha_2|\phi_a\rangle-\alpha_1|\phi_b\rangle}{\beta_2}.
\end{equation}
By construction, $\langle 2_{\mathrm D}|2_{\mathrm K}\rangle=0$, and the pair $\{|2\rangle_{\mathrm K},|2\rangle_{\mathrm D}\}$ spans the same two-dimensional occupation subspace. Using that $\hat H_{\mathrm{rot}}|\phi_a\rangle=4B|\phi_a\rangle$ and $\hat H_{\mathrm{rot}}|\phi_b\rangle=6B|\phi_b\rangle$, the $2\times 2$ matrix elements are:
\begin{align}
\langle 2_{\mathrm K}|\hat H_{\mathrm{rot}}|2_{\mathrm K}\rangle &= \frac{4B|\alpha_1|^2+6B|\alpha_2|^2}{|\alpha_1|^2+|\alpha_2|^2}=E_2^{\mathrm{(br)}},\\
\langle 2_{\mathrm D}|\hat H_{\mathrm{rot}}|2_{\mathrm D}\rangle &= \frac{4B|\alpha_2|^2+6B|\alpha_1|^2}{|\alpha_1|^2+|\alpha_2|^2},\\
\langle 2_{\mathrm D}|\hat H_{\mathrm{rot}}|2_{\mathrm K}\rangle &= \frac{(4B-6B)\alpha_1 \alpha_2}{|\alpha_1|^2+|\alpha_2|^2} = -\frac{2B\,\alpha_1 \alpha_2}{|\alpha_1|^2+|\alpha_2|^2}\neq 0.
\end{align}
Therefore, the Krylov bright state $|2\rangle_{\mathrm K}$ is not an eigenstate of $\hat H_{\mathrm{rot}}$ and we leak into the dark state $|2\rangle_{\mathrm D}$ whenever the single-rotor ladder is anharmonic, $E_2\neq 2E_1$. In the large $N$ limit, the diagonal matrix elements $\langle 2_{\mathrm K}|\hat H_{\mathrm{rot}}|2_{\mathrm K}\rangle$ and $\langle 2_{\mathrm D}|\hat H_{\mathrm{rot}}|2_{\mathrm D}\rangle$ tend to a constant, whereas the off-diagonal $\langle 2_{\mathrm D}|\hat H_{\mathrm{rot}}|2_{\mathrm K}\rangle$ scales as $\mathcal{O} (1/\sqrt{N})$, indicating that the dark-state leakage matrix element is suppressed for large $N$. We note that this suppression reflects that $|2\rangle_{\mathrm K}$ is dominated by the many-rotor configuration with two $J=1$ excitations, $|\phi_a\rangle$, whose amplitude scales as $\sqrt{N}$, while the configuration with one $J=2$ excitation carries only an $O(1)$ amplitude.

\newpage

\subsection{Schrieffer--Wolff transformation up to fourth order}
We start from the collisional light--matter Hamiltonian
\begin{equation}
\label{eq:Hcoll}
\hat H_{\mathrm{coll}}
=
\omega_c\,\hat a^\dagger \hat a
+\sum_{L\ge 0} E_L\,|L\rangle\langle L|
-g_0(\hat a+\hat a^\dagger)\sum_{L\ge 0}c_{L+1}
\Big(|L\rangle\langle L{+}1|+\mathrm{h.c.}\Big),
\end{equation}
with real couplings $c_{L+1}$ and $c_0=0$. Truncating to three rotor
levels $\{|0\rangle,|1\rangle,|2\rangle\}$ gives
\begin{equation}
\label{eq:Hcoll_3lvl}
\hat H_{\mathrm{coll}}^{(3)}
=
\omega_c\,\hat a^\dagger \hat a
+\sum_{j=0}^2 E_j\,|j\rangle\langle j|
-g_0(\hat a+\hat a^\dagger)\Big[
c_{01}\big(|0\rangle\langle 1|+\mathrm{h.c.}\big)
+c_{12}\big(|1\rangle\langle 2|+\mathrm{h.c.}\big)
\Big],
\end{equation}
where $c_{01}\equiv c_1$ and $c_{12}\equiv c_2$.
We construct the Schrieffer--Wolff (SW) block diagonalization up to
$\mathcal O(g_0^4)$ and then project onto the rotor ground manifold
$L=0$. Let
\begin{equation}
\omega_{01}\equiv E_1-E_0,
\qquad
\omega_{02}\equiv E_2-E_0.
\end{equation}
and further the detunings
\begin{equation}
\label{eq:new_detunings}
d_{01}\equiv \omega_{01}-\omega_c,
\qquad
d_{02}\equiv \omega_{02}-2\omega_c,
\end{equation}
together with the shorthand couplings
\begin{equation}
g_{01}\equiv g_0 c_{01},
\qquad
g_{12}\equiv g_0 c_{12}.
\end{equation}
Thus
\begin{align}
d_{01}+2\omega_c &= \omega_{01}+\omega_c, &
d_{01}+4\omega_c &= \omega_{01}+3\omega_c, &
-d_{01}+2\omega_c &= 3\omega_c-\omega_{01},
\\
d_{02}+2\omega_c &= \omega_{02}, &
d_{02}+4\omega_c &= \omega_{02}+2\omega_c .
\end{align}
Let
\begin{equation}
P\equiv |0\rangle\langle 0|\otimes\mathbb I_{\rm cav}
\end{equation}
be the projector onto the rotor ground manifold. The SW-transformed
Hamiltonian
\begin{equation}
\hat H' = e^{\hat S}\hat H_{\rm coll}^{(3)}e^{-\hat S},
\qquad
\hat S^\dagger=-\hat S,
\end{equation}
is chosen such that it is block diagonal up to $\mathcal O(g_0^4)$.
Projecting to $L=0$ yields the effective cavity Hamiltonian
\begin{equation}
\label{eq:Heff_form}
\hat H_{\rm eff}^{(L=0)}
\equiv P\hat H'P
=
E_0^{\rm (eff)}
+ A(\hat a^{\dagger 2}+\hat a^2)
+ \Omega\,\hat a^\dagger \hat a
+ \kappa(\hat a^{\dagger 4}+\hat a^4)
+ \mu(\hat a^{\dagger 3}\hat a+\hat a^\dagger \hat a^3)
+ \nu\,\hat a^{\dagger 2}\hat a^2
+\mathcal O(g_0^6).
\end{equation}
The coefficients below are written without making a rotating-wave
approximation using a CAS to calculate the commutators analytically. Notably, all but $\nu$ arise already for finite $g_{01}$, while $\nu$ explicitely requires $|g_{12}| >0$ and makes necessary a multi-level system. The ground-state energy is
\begin{equation}
E_0^{\rm (eff)}=E_0+\mathcal E_0 .
\end{equation}
with
\begin{equation}
\label{eq:E0_full_nonrwa}
\mathcal E_0
=
-\frac{
g_{01}^2
\Big[
d_{01}d_{02}^2
+6d_{01}d_{02}\omega_c
+8d_{01}\omega_c^2
+2d_{02}^2\omega_c
+3d_{02}g_{12}^2
+12d_{02}\omega_c^2
+8g_{12}^2\omega_c
+16\omega_c^3
\Big]
}{
(d_{01}+2\omega_c)^2(d_{02}+2\omega_c)(d_{02}+4\omega_c)
}.
\end{equation}
The renormalized cavity frequency is
\begin{align}
\label{eq:Omega_full_nonrwa}
\Omega
&=
\frac{1}{
d_{01}^2(d_{01}+2\omega_c)^2(d_{02}+2\omega_c)(d_{02}+4\omega_c)
}
\Big[
d_{01}^4d_{02}^2\omega_c
+6d_{01}^4d_{02}\omega_c^2
+8d_{01}^4\omega_c^3
\nonumber\\
&\quad
-2d_{01}^3d_{02}^2g_{01}^2
+4d_{01}^3d_{02}^2\omega_c^2
-12d_{01}^3d_{02}g_{01}^2\omega_c
+24d_{01}^3d_{02}\omega_c^3
-16d_{01}^3g_{01}^2\omega_c^2
+32d_{01}^3\omega_c^4
\nonumber\\
&\quad
-6d_{01}^2d_{02}^2g_{01}^2\omega_c
+4d_{01}^2d_{02}^2\omega_c^3
-12d_{01}^2d_{02}g_{01}^2g_{12}^2
-36d_{01}^2d_{02}g_{01}^2\omega_c^2
+24d_{01}^2d_{02}\omega_c^4
\nonumber\\
&\quad
-40d_{01}^2g_{01}^2g_{12}^2\omega_c
-48d_{01}^2g_{01}^2\omega_c^3
+32d_{01}^2\omega_c^5
-4d_{01}d_{02}^2g_{01}^2\omega_c^2
-12d_{01}d_{02}g_{01}^2g_{12}^2\omega_c
\nonumber\\
&\quad
-24d_{01}d_{02}g_{01}^2\omega_c^3
-48d_{01}g_{01}^2g_{12}^2\omega_c^2
-32d_{01}g_{01}^2\omega_c^4
-4d_{02}g_{01}^2g_{12}^2\omega_c^2
-16g_{01}^2g_{12}^2\omega_c^3
\Big].
\end{align}

The two-photon squeezing coefficient is
\begin{align}
\label{eq:A_full_nonrwa}
A
&=
-\frac{g_{01}^2}{
d_{01}^3d_{02}(d_{01}+2\omega_c)^3(d_{01}+4\omega_c)
(d_{02}+2\omega_c)(d_{02}+4\omega_c)
}
\Big[
d_{01}^6d_{02}^3
+6d_{01}^6d_{02}^2\omega_c
+8d_{01}^6d_{02}\omega_c^2
\nonumber\\
&\quad
+9d_{01}^5d_{02}^3\omega_c
+6d_{01}^5d_{02}^2g_{12}^2
+54d_{01}^5d_{02}^2\omega_c^2
+20d_{01}^5d_{02}g_{12}^2\omega_c
+72d_{01}^5d_{02}\omega_c^3
+4d_{01}^5g_{12}^2\omega_c^2
\nonumber\\
&\quad
+28d_{01}^4d_{02}^3\omega_c^2
+42d_{01}^4d_{02}^2g_{12}^2\omega_c
+168d_{01}^4d_{02}^2\omega_c^3
+154d_{01}^4d_{02}g_{12}^2\omega_c^2
+224d_{01}^4d_{02}\omega_c^4
+40d_{01}^4g_{12}^2\omega_c^3
\nonumber\\
&\quad
+36d_{01}^3d_{02}^3\omega_c^3
+102d_{01}^3d_{02}^2g_{12}^2\omega_c^2
+216d_{01}^3d_{02}^2\omega_c^4
+416d_{01}^3d_{02}g_{12}^2\omega_c^3
+288d_{01}^3d_{02}\omega_c^5
+144d_{01}^3g_{12}^2\omega_c^4
\nonumber\\
&\quad
-24d_{01}^2d_{02}^3g_{01}^2\omega_c^2
+16d_{01}^2d_{02}^3\omega_c^4
-144d_{01}^2d_{02}^2g_{01}^2\omega_c^3
+108d_{01}^2d_{02}^2g_{12}^2\omega_c^3
+96d_{01}^2d_{02}^2\omega_c^5
\nonumber\\
&\quad
-192d_{01}^2d_{02}g_{01}^2\omega_c^4
+488d_{01}^2d_{02}g_{12}^2\omega_c^4
+128d_{01}^2d_{02}\omega_c^6
+224d_{01}^2g_{12}^2\omega_c^5
\nonumber\\
&\quad
-56d_{01}d_{02}^3g_{01}^2\omega_c^3
-336d_{01}d_{02}^2g_{01}^2\omega_c^4
+48d_{01}d_{02}^2g_{12}^2\omega_c^4
-448d_{01}d_{02}g_{01}^2\omega_c^5
+224d_{01}d_{02}g_{12}^2\omega_c^5
+128d_{01}g_{12}^2\omega_c^6
\nonumber\\
&\quad
-32d_{02}^3g_{01}^2\omega_c^4
-192d_{02}^2g_{01}^2\omega_c^5
-256d_{02}g_{01}^2\omega_c^6
\Big].
\end{align}

The four-photon coefficient is
\begin{align}
\label{eq:kappa_full_nonrwa}
\kappa
&=
\frac{
g_{01}^2
}{
d_{01}^2d_{02}(-d_{01}+2\omega_c)(d_{01}+2\omega_c)^2
(d_{01}+4\omega_c)(d_{02}+4\omega_c)
}
\Big[
d_{01}^4d_{02}g_{12}^2
+2d_{01}^4g_{12}^2\omega_c
\nonumber\\
&\quad
+4d_{01}^3d_{02}g_{12}^2\omega_c
+16d_{01}^3g_{12}^2\omega_c^2
+8d_{01}^2d_{02}g_{12}^2\omega_c^2
+40d_{01}^2g_{12}^2\omega_c^3
\nonumber\\
&\quad
-16d_{01}d_{02}^2g_{01}^2\omega_c^2
-64d_{01}d_{02}g_{01}^2\omega_c^3
+8d_{01}d_{02}g_{12}^2\omega_c^3
+32d_{01}g_{12}^2\omega_c^4
\nonumber\\
&\quad
-16d_{02}^2g_{01}^2\omega_c^3
-64d_{02}g_{01}^2\omega_c^4
\Big].
\end{align}
The coefficient of
$\hat a^{\dagger 3}\hat a+\hat a^\dagger\hat a^3$ is
\begin{align}
\label{eq:mu_full_nonrwa}
\mu
&=
-\frac{
4g_{01}^2
}{
d_{01}^2d_{02}(-d_{01}+2\omega_c)(d_{01}+2\omega_c)^2
(d_{01}+4\omega_c)(d_{02}+2\omega_c)(d_{02}+4\omega_c)
}
\Big[
-d_{01}^4d_{02}^2g_{12}^2
\nonumber\\
&\quad
-4d_{01}^4d_{02}g_{12}^2\omega_c
-2d_{01}^4g_{12}^2\omega_c^2
-4d_{01}^3d_{02}^2g_{12}^2\omega_c
-19d_{01}^3d_{02}g_{12}^2\omega_c^2
-14d_{01}^3g_{12}^2\omega_c^3
\nonumber\\
&\quad
-9d_{01}^2d_{02}g_{12}^2\omega_c^3
-24d_{01}^2g_{12}^2\omega_c^4
+4d_{01}d_{02}^3g_{01}^2\omega_c^2
+24d_{01}d_{02}^2g_{01}^2\omega_c^3
+8d_{01}d_{02}^2g_{12}^2\omega_c^3
\nonumber\\
&\quad
+32d_{01}d_{02}g_{01}^2\omega_c^4
+34d_{01}d_{02}g_{12}^2\omega_c^4
+8d_{01}g_{12}^2\omega_c^5
+4d_{02}^3g_{01}^2\omega_c^3
+24d_{02}^2g_{01}^2\omega_c^4
\nonumber\\
&\quad
+8d_{02}^2g_{12}^2\omega_c^4
+32d_{02}g_{01}^2\omega_c^5
+40d_{02}g_{12}^2\omega_c^5
+32g_{12}^2\omega_c^6
\Big].
\end{align}
Finally, the Kerr-type normal-ordered quartic coefficient is
\begin{align}
\label{eq:nu_full_nonrwa}
\nu
&=
-\frac{
2g_{01}^2g_{12}^2
}{
d_{01}^2d_{02}(d_{01}+2\omega_c)^2(d_{02}+2\omega_c)(d_{02}+4\omega_c)
}
\Big[
3d_{01}^2d_{02}^2
+12d_{01}^2d_{02}\omega_c
+4d_{01}^2\omega_c^2
\nonumber\\
&\quad
+6d_{01}d_{02}^2\omega_c
+28d_{01}d_{02}\omega_c^2
+16d_{01}\omega_c^3
+4d_{02}^2\omega_c^2
+20d_{02}\omega_c^3
+16\omega_c^4
\Big].
\end{align}

\end{widetext}

\end{document}